\documentclass[10pt]{iopart}
\usepackage{iopams}  
\usepackage{epsfig}
\usepackage{graphicx}

\newcommand{\bea}{\begin{eqnarray*}}
\newcommand{\eea}{\end{eqnarray*}}

\newcommand{\deltacp}{\delta_\mathrm{CP}}

\newcommand{\dm}[1]{{\Delta m^2_{#1}}}

\newcommand{\ie}{{\it i.e.}}

\newcommand{\JHFHK}{JHF-HK}
\newcommand{\JHFSK}{JHF-SK}
\newcommand{\NuFactI}{NuFact-I}
\newcommand{\NuFactII}{NuFact-II}

\begin{document}

\vspace*{-3.cm}
\begin{flushright}
TUM-HEP-480/02\\
MPI-PhT/2000-53\\
\end{flushright}

\vspace*{0.5cm}

\title{CP, T and CPT violation in future long baseline experiments}

\author{Patrick Huber} 

\address{Institut f{\"u}r theoretische Physik, Physik  Department, Technische Universi{\"a}t M{\"u}nchen,
James-Franck-Strasse, D-85748 Garching}

\address{Max-Planck Institut f{\"u}r Physik, Postfach 401212, D-80805 
M{\"u}nchen}

\begin{abstract}
I give a short overview about the possibilities and problems related to
the measurement of CP violation in long baseline experiments. Special 
attention is paid to the issue of degeneracies and a method for their 
resolution is quantitatively discussed. The CP violation reach for different 
experiments is compared in dependence of $\sin^22\theta_{13}$ and $\dm{21}$. 
Furthermore a short comment  about the possible effects of matter induced
T violation is made. Finally the limits on CPT violation obtainable at a 
neutrino factory are shown.

\end{abstract}

\section{CP violation}
In contrast to the quark sector CP violation in the lepton sector can be 
potentially large. This has spurred great interest in the possible measurement
of the leptonic Dirac-type CP-phase $\delta_{CP}$ especially in the context of
a planed neutrino factory~\cite{DeRujula:1998hd,Yasuda:2001ip}. The principal
observable is the CP-odd probability difference $\Delta P_{\alpha\beta}^{CP}
=P_{(\alpha\rightarrow\beta)}-P_{(\bar{\alpha}\rightarrow\bar{\beta})}$.
Since the neutrinos travel a long distance trough the Earth matter one has to
include matter effects. The Earth is however CP-asymmetric by itself, thus it
introduces a non-vanishing $\Delta P_{\alpha\beta}^{CP}$ even if 
$\delta_{CP}=0$. This makes it difficult to use $\Delta P_{\alpha\beta}^{CP}$ 
as measure of  $\delta_{CP}$. Furthermore there are, due to the form of 
the oscillation  probabilities, strong correlations among several oscillation 
parameters. Besides that a long baseline experiment does not measure the 
probabilities themselves but event rates. There are many systematical 
uncertainties in translating a rate measurement into a measurement of the 
probability. In addition to those 
problems the oscillation probabilities allow for different sets of parameters 
which give approximately the same probabilities. This is known as the 
degeneracy problem. In observing only the transition between electron and muon neutrinos or anti-neutrinos there remain three possible degeneracies:
the $(\delta_{CP},\theta_{13})$ ambiguity~\cite{Burguet-Castell:2001ez}, the
$\mathrm{sign} \dm{31}$ degeneracy~\cite{Minakata:2001qm} and the $(\theta_{23},\pi/4 - \theta_{23})$ degeneracy~\cite{BARGER}. Those degeneracies can have a
substantial impact on the ability of a given experiment to reach its physics 
goals. The results shown in the following are a small subset of the
results obtained in~\cite{Huber:2002mx}.

The setups considered are listed in table~\ref{tab:setups}. Further details 
can be found in~\cite{Huber:2002mx}.
\begin{table}[h!]
\begin{flushright}
\begin{tabular}{c|c|c|c}
\JHFSK&\JHFHK&\NuFactI&\NuFactII\\
$22.5\,\mathrm{kt}$&$1\,000\,\mathrm{kt}$&$10\,\mathrm{kt}$&$50\,\mathrm{kt}$\\
water Cherenkov&water Cherenkov&magnetized iron& magnetized iron\\
&&calorimeter&calorimeter\\
$0.75\,\mathrm{MW}$&$4\,\mathrm{MW}$&$0.75\,\mathrm{MW}$&$4\,\mathrm{MW}$\\
5 years&8 years&5 years&8 years\\
\end{tabular}
\end{flushright}
\caption{\label{tab:setups} Definition of experimental setups.}
\end{table}
Of all the systematical errors considered in~\cite{Huber:2002mx} the most
important for the JHF setups is the background normalization uncertainty,
whereas the NuFact setups are in general little hindered by systematical 
errors.

In~\cite{Huber:2002mx} a complete analysis of the multi parameter correlations
was performed, taking into account external information on $\dm{21}$ provided 
by KamLand and on the matter density provided by geophysics. For $\dm{21}$ an 
error of $15\%$ and for the matter density an error of $5\%$ is assumed. 
The biggest source of correlation errors for all setups is the correlation 
between $\theta_{13}$ and $\delta_{CP}$. This is an intrinsic effect which is
very hard to fight. For the NuFact scenarios the matter density plays a crucial
rule and one should seek to improve the knowledge on this quantity.

\begin{figure}[ht!]
\begin{flushright}
\includegraphics[angle=-90,width=12cm]{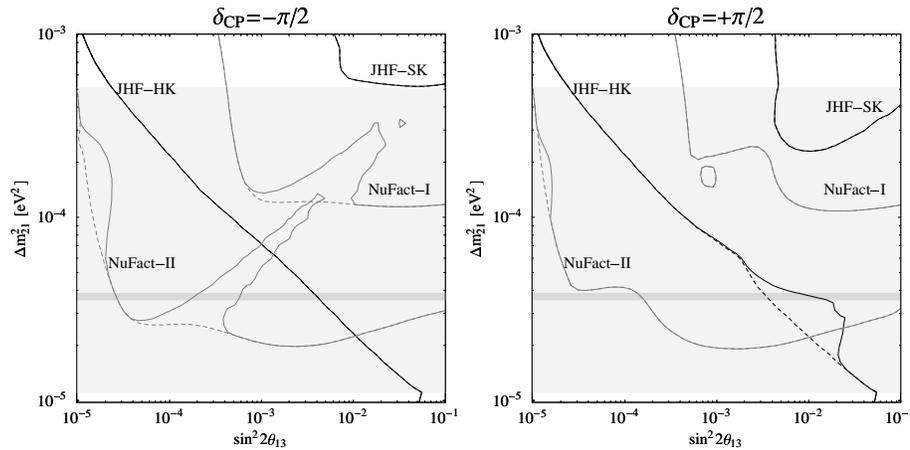}
\caption{\label{fig:cpv}The sensitivity to CP violation for all 
experiments (as labeled in the plots) at the $2 \sigma$
confidence level, plotted as function of $\sin^2 2 \theta_{13}$ and $\Delta
m_{21}^2$. The left--hand plot shows the case of $\deltacp = - \pi/2$ and the
right--hand plot the case of $\deltacp = + \pi/2$. Solid curves refer to taking
into account all degenerate solutions and dashed curves to taking into
account the best--fit manifold only. The gray shaded regions refer to the
allowed LMA region for $\Delta m_{21}^2$ (light gray) and the best--fit value
for $\Delta m_{21}^2$ (dark gray).} 
\end{flushright}
\end{figure}

The $(\theta_{23},\pi/4 - \theta_{23})$ degeneracy has only little impact on
the determination of $\delta_{CP}$ in all cases. The influence of the 
$(\delta_{CP},\theta_{13})$ ambiguity is strongest at a NuFact and its effects
strongly depend on subtle details of the expected detector performance, a 
detailed discussion is given in appendix B of~\cite{Huber:2002mx}. The
$\mathrm{sign} \dm{31}$ degeneracy however has a substantial effect on the
ability to determine $\delta_{CP}$, especially at a NuFact it may be rather 
cumbersome. In figure~\ref{fig:cpv} the CP violation reach in the 
$\sin^22\theta_{13}$-$\dm{21}$ plane is shown for all four setups including
degeneracies, correlations, systematics and backgrounds.
The trench in the left hand panel for the {\NuFactII} case is due to the
$\mathrm{sign} \dm{31}$ degeneracy.
One possibility among many others~\cite{Donini:2002rm,Burguet-Castell:2002qx} 
to resolve the correlation between $\delta_{CP}$ and 
$\theta_{13}$ and to break the $(\delta_{CP},\theta_{13})$ ambiguity is
to use the so called ``magic baseline''. The condition for the magic baseline
is given by $\sin (\dm{31} L/(4E)\quad 2VE/\dm{31})=0 \Leftrightarrow VL/2=n
\pi$~\cite{BARGER}, this gives for Earth densities a baseline 
$L\simeq 8\,100\,\mathrm{km}$. At this special distance all terms in the 
appearance probability proportional to $\dm{21}$ and $(\dm{21})^2$ vanish
identically for all energies and values of $\dm{31}$, {\ie} the appearance 
probability reduces to the one in a two neutrino case. Thus all correlations
and degeneracies connected to $\delta_{CP}$ disappear. 
\begin{figure}[ht!]
\begin{center}
\includegraphics[width=5cm]{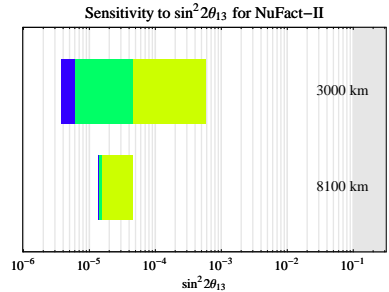}
\caption{\label{fig:magic} The sensitivity to $\sin^22\theta_{13}$ at baselines
of $3\,000\,\mathrm{km}$ and $8\,100\,\mathrm{km}$. The dark grey shading 
indicates the impact of systematics, the medium grey shading the impact 
of the correlations and the light grey shading stands for the effect of 
degeneracies.}
\end{center}
\end{figure}
The drawback of this
very long baseline is that the event rates decrease. But in 
figure~\ref{fig:magic} it is clearly visible that the gain in sensitivity
by avoiding the effects of $\delta_{CP}$ (rightmost edge) is much larger 
than the effects of the diminished statistics (leftmost edge), thus the
sensitivity is increased by one order of magnitude.

\section{T and CPT violation}

In vacuum the CP-odd and T-odd probability differences are identical. This
does not hold in matter. In principle it is
therefore possible that matter profile asymmetries introduce a fake CP 
violation and increase the error in the measurement of $\delta_{CP}$. However 
in~\cite{Akhmedov:2001kd} it is shown that, for in terrestial experiments
conceivable matter asymmetries, this effects turns out to be negligibly small.

CPT violation would manifest itself in a neutrino oscillation experiment by
the presence of two different $\dm{}$ scales or mixing angles for neutrinos 
and anti-neutrinos. Thus in order to estimate the sensitivity of a neutrino 
factory to CPT violation one just needs to evaluate the level of accuracy
which can be obtained in the measurement of $\dm{31}$ and $\theta_{23}$. 
At a standard NuFact relative asymmetries in the mass splittings of order
$<10^{-1}$ and in the mixing angles in the order of $<10^{-2}$ could be 
detected, as it is shown in~\cite{Bilenky:2001ka}.


\vspace{0.5cm}
\begin{thebibliography}{10}

\expandafter\ifx\csname bibnamefont\endcsname\relax
  \def\bibnamefont#1{#1}\fi
\expandafter\ifx\csname bibfnamefont\endcsname\relax
  \def\bibfnamefont#1{#1}\fi
\expandafter\ifx\csname url\endcsname\relax
  \def\url#1{\texttt{#1}}\fi
\expandafter\ifx\csname urlprefix\endcsname\relax\def\urlprefix{URL }\fi
\providecommand{\bibinfo}[2]{#2}
\providecommand{\eprint}[2][]{\url{#2}}

\bibitem{DeRujula:1998hd}
\bibinfo{author}{\bibfnamefont{A.~D.} \bibnamefont{Rujula}},
  \bibinfo{author}{\bibfnamefont{M.~B.} \bibnamefont{Gavela}},
  \bibnamefont{and}
  \bibinfo{author}{\bibfnamefont{P.}~\bibnamefont{Hernandez}},
  \bibinfo{journal}{Nucl. Phys.} \textbf{\bibinfo{volume}{B547}},
  \bibinfo{pages}{21} (\bibinfo{year}{1999}), \eprint{hep-ph/9811390}.

\bibitem{Yasuda:2001ip}
O.~Yasuda,
\eprint{hep-ph/0111172}, and references therein.

\bibitem{Burguet-Castell:2001ez}
\bibinfo{author}{\bibfnamefont{J.}~\bibnamefont{Burguet-Castell}}
  {\it et al.},
  \bibinfo{journal}{Nucl. Phys.} \textbf{\bibinfo{volume}{B608}},
  \bibinfo{pages}{301} (\bibinfo{year}{2001}),
  \eprint[http://arXiv.org/abs]{hep-ph/0103258}.

\bibitem{Minakata:2001qm}
\bibinfo{author}{\bibfnamefont{H.}~\bibnamefont{Minakata}} \bibnamefont{and}
  \bibinfo{author}{\bibfnamefont{H.}~\bibnamefont{Nunokawa}},
  \bibinfo{journal}{JHEP} \textbf{\bibinfo{volume}{10}}, \bibinfo{pages}{001}
  (\bibinfo{year}{2001}), \eprint[http://arXiv.org/abs]{hep-ph/0108085}.

\bibitem{BARGER}
\bibinfo{author}{\bibfnamefont{V.}~\bibnamefont{Barger}},
  \bibinfo{author}{\bibfnamefont{D.}~\bibnamefont{Marfatia}}, \bibnamefont{and}
  \bibinfo{author}{\bibfnamefont{B.}~\bibnamefont{Wood}},
  \bibinfo{journal}{Phys. Lett.} \textbf{\bibinfo{volume}{B498}},
  \bibinfo{pages}{53} (\bibinfo{year}{2001}), \eprint{hep-ph/0011251}.

\bibitem{Huber:2002mx}
P.~Huber, M.~Lindner and W.~Winter,
\eprint[http://arXiv.org/abs]{hep-ph/0204352}, to appear in Nucl. Phys. 
{\bf B}.

\bibitem{Donini:2002rm}
A.~Donini, D.~Meloni and P.~Migliozzi,
\eprint{hep-ph/0206034}, see also this proceedings.

\bibitem{Burguet-Castell:2002qx}
J.~Burguet-Castell {\it et al.},
\eprint{hep-ph/0207080}, see also this proceedings.

\bibitem{Akhmedov:2001kd}
E.~K.~Akhmedov {\it et al.}, 
Nucl.\ Phys. {\bf B608} (2001) 394,
\eprint[http://arXiv.org/abs]{hep-ph/0105029}.

\bibitem{Bilenky:2001ka}
S.~M.~Bilenky {\it et al.}, 
Phys.\ Rev. {\bf D65} (2002) 073024,
\eprint{hep-ph/0112226}.




\end{thebibliography}
\end{document}